\documentclass[fleqn,usenatbib]{mnras}

\usepackage{newtxtext,newtxmath}

\usepackage[T1]{fontenc}

\DeclareRobustCommand{\VAN}[3]{#2}
\let\VANthebibliography\thebibliography
\def\thebibliography{\DeclareRobustCommand{\VAN}[3]{##3}\VANthebibliography}

\newcommand{\edit}[1]{{#1}}

\usepackage{graphicx}	
\usepackage{amsmath}	

\title[The mass-size relation of galaxy clusters]{The mass-size relation of galaxy clusters}

\author[O. Contigiani et al.]{
O. Contigiani,$^{1,2}$\thanks{E-mail:contigiani@strw.leidenuniv.nl}
Y. M. Bah\'e,$^{1}$
H. Hoekstra$^{1}$
\\
$^{1}$Leiden Observatory, Leiden University, PO Box 9506, Leiden 2300 RA, The Netherlands \\
$^{2}$Lorentz Institute for Theoretical Physics, Leiden University, PO Box 9506, Leiden 2300 RA, The Netherlands
}

\date{Accepted XXX. Received YYY; in original form ZZZ}

\pubyear{2015}

\begin{document}
\label{firstpage}
\pagerange{\pageref{firstpage}--\pageref{lastpage}}
\maketitle

\begin{abstract}
The outskirts of accreting dark matter haloes exhibit a sudden drop in density delimiting their multi-stream region. Due to the dynamics of accretion, the location of this physically motivated edge strongly correlates with the halo growth rate. Using hydrodynamical zoom-in simulations of high-mass clusters, we explore this definition in realistic simulations and find an explicit connection between this feature in the dark matter and galaxy profiles. We also show that the depth of the splashback feature correlates well with the direction of filaments and, surprisingly, the orientation of the brightest cluster galaxy. Our findings suggest that galaxy profiles and weak-lensing masses can define an observationally viable mass-size scaling relation for galaxy clusters, which can be used to extract cosmological information.  
\end{abstract}




\section{Introduction}

In the $\Lambda$CDM paradigm, structure in the Universe arises from the initial density perturbations of an (almost) homogeneous dark matter distribution. Due to gravitational evolution, this leads to the appearance of collapsed structures, i.e. dark matter haloes. Some of the baryonic matter, following this process, cools down and settles at the centres of the gravitational potentials where it forms galaxies.

This mechanism has been studied through models of so-called spherical collapse \citep{Gunn1972, 1985ApJS...58...39B}, whose main prediction is the existence of a radius within which the material orbiting the halo is completely virialized. In general, this virial radius depends on cosmology and redshift, but both in numerical simulations and observations, fixed overdensity radii are widely used as proxies for this quantity. An example of this is $r_{200m}$, defined as the radius within which the average density is $200$ times the average matter density of the Universe, $\rho_\text{m}$. The corresponding enclosed mass is known as $M_{200m}$.

Halo mass functions constructed with these idealized definitions can capture the effects of cosmology \citep{1974ApJ...187..425P}, the nature of dark matter \citep{2013MNRAS.434.3337A}, and dark energy \citep{Mead_2016} on the growth of structure. In the real Universe, however, this picture is complicated by the triaxiality of haloes \citep{1991ApJ...378..496D, Monaco_1995} and the existence of clumpy (baryonic) substructure \citep{Bocquet_2015}.

Because the process of structure formation is hierarchical, massive haloes contain subhaloes, some of which host galaxies themselves. The resulting clusters of galaxies are the focus of this work. What makes these objects particularly unique is the fact that they are not fully virialized yet. To this day, they are still accreting both ambient material and subhaloes through filamentary structures surrounding them \citep{Bond_1996}. Because of their definition, however, traditional overdensity definitions of mass are not only affected by halo growth, but also by a pseudo-evolution due to  the redshift dependence of $\rho_\text{m}$ \citep{2013ApJ...766...25D}.

\cite{Diemer_2014} and \cite{More_2015} were the first to note that this growth process leads to the formation of a sharp feature in the density profile that separates collapsed and infalling material. This feature, therefore, defines a natural boundary of the halo. The location of this edge, i.e. the splashback radius $r_\text{sp}$, has an obvious primary dependence on halo mass, but also a secondary dependence on accretion rate. While this behaviour can be qualitatively explained using simple semi-analytical models of spherical collapse, none of the analytical models currently proposed \citep{Adhikari_2014, Shi2016} can fully describe its dependency on mass and accretion rate \citep{Diemer_2017b}. Despite this, the corresponding definition of halo mass is particularly suited to define a universal mass function valid for a wide range of cosmologies  \citep{diemer2020universal}. 

In this paper, we try to bridge the gap between the theoretical understanding of the splashback feature and observational results, both past and future. The outer edge of clusters has already been extensively measured through different tracers: the radial distribution of galaxies from wide surveys \citep{2016ApJ...825...39M, Baxter_2017, Chang_2018}, but also their velocity distribution \citep{tomooka2020clusters, 2021MNRAS.503.4250F}, and in the weak-lensing signal of massive clusters \citep{Umetsu_2017, Chang_2018, Contigiani_2019b}. Furthermore, forecasts have already set expectations for what will be obtainable from near-future experiments \citep{Fong_2018, xhakaj2019accurately, 2020arXiv201011324W}. Despite the wealth of data and studies, however, not many \emph{splashback observables} have been proposed. The only robust application of this feature found in the literature is related to the study of quenching for newly accreted galaxies \citep{2020arXiv200811663A}.

To achieve our goal, we make use of hydrodynamical simulations of massive galaxy clusters, which we introduce in Section~\ref{sec:Hydrangea}. We focus mainly on $z=0$, but also make use of snapshots at redshifts $z=0.474$ and $z=1.017$. In Section \ref{sec:definition}, we start our discussion by introducing the physical interpretation of splashback and consider the connection between the galaxy and dark matter distributions. We then continue in Section \ref{sec:msrelation} and \ref{sec:redshift}, where we explain how galaxy profiles and weak-lensing mass measurements can be combined to construct a mass-size relationship for galaxy clusters. Finally, we summarize our conclusions and suggest future developments in Section \ref{sec:conclusions}.

\section{Hydrangea}
\label{sec:Hydrangea}

The Hydrangea simulations are a suite of $24$ zoom-in hydrodynamical simulations of massive galaxy clusters ($\log_{10}M_\text{200m}/\mathrm{M}_\odot$ between $14$ and $15.5$ at redshift $z = 0$) designed to study the relationship between galaxy formation and cluster environment \citep{Bahe2017}. They are part of the Cluster-EAGLE project \citep{Bahe2017,Barnes_2017} and have been run using the EAGLE galaxy formation model \citep{Schaye_2014}, which is known to reproduce galaxy observables such as colour distribution and star formation rates. To better reproduce the observed hot gas fractions in galaxy groups, the AGNdT9 variant of this model was used \citep{Schaye_2014}. 

The zoom-in regions stretch to between 10 and 30 Mpc from the cluster centre, roughly corresponding to $\lesssim 10 r_{200m}$. For the definition of the cluster centre, in this work, we choose the minimum of the gravitational potential. We note, however, that this choice will not impact our conclusions since we will focus on locations around $r_\text{200m}$. The particle mass of $m \sim 10^{6}~\mathrm{M}_\odot$ for baryons and $m\sim10^{7}~\mathrm{M}_\odot$ for dark matter allows us to resolve galaxy positions down to stellar masses $M_\ast \geq 10^{8}~ \mathrm{M}_\odot$ and total masses $M_\text{sub} \geq 10^{9}~
\mathrm{M}_\odot$, respectively. 

\begin{figure}
    \centering
    \includegraphics[width=0.46\textwidth]{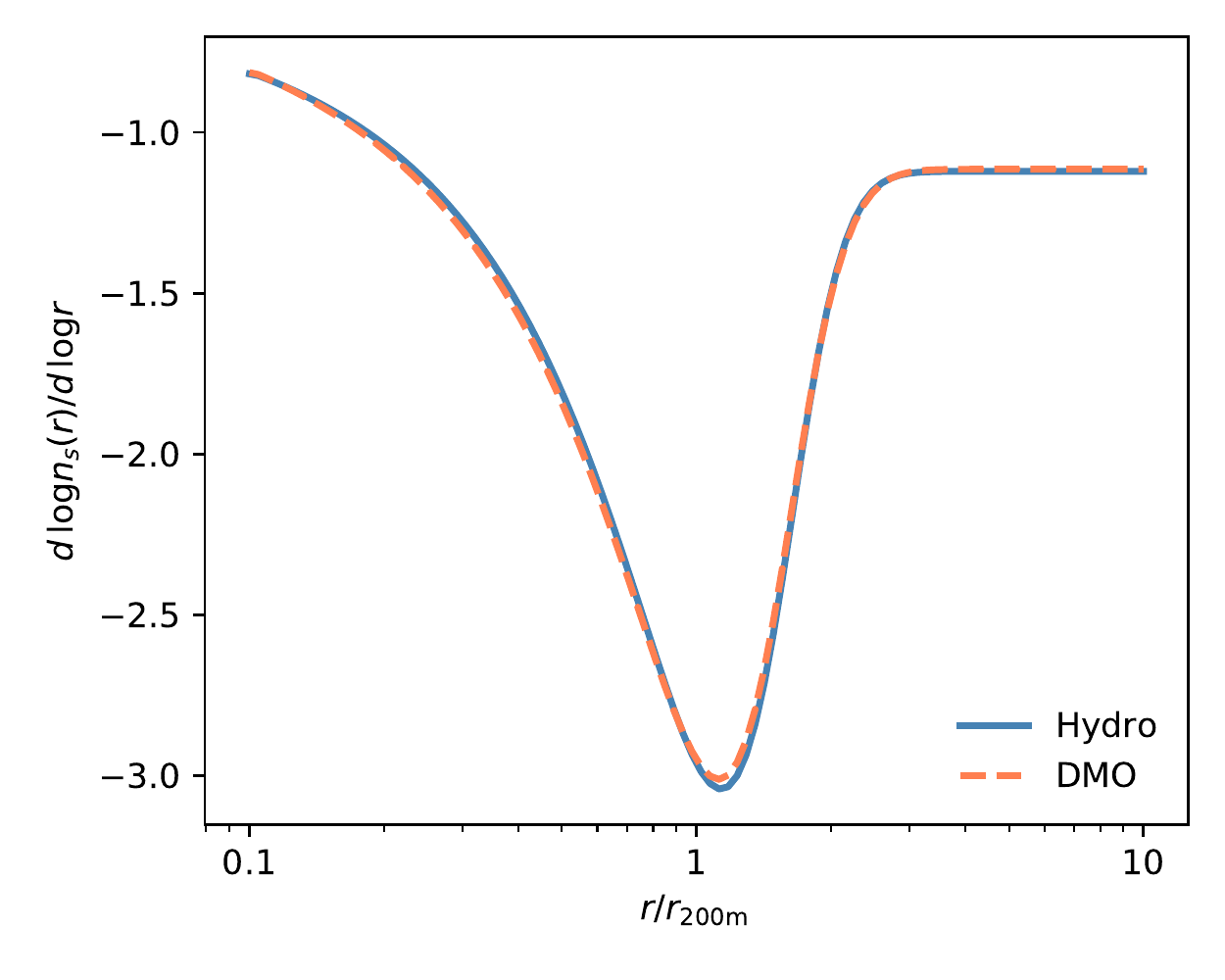}
    \caption{The splashback feature visible in the average subhalo distribution of simulated high-mass clusters. We extract the logarithmic slope by fitting a smooth profile to the mean of the Hydrangea profiles rescaled by $r_\text{200m}$. We perform this operation both on the hydrodynamical simulations (Hydro) and their dark matter only counterparts (DMO). The minimum around $r_{200m}$ marks the halo boundary, and this figure highlights the lack of baryonic effects on the location or depth of this feature. The two logarithmic slope profiles are consistent with each other at the $1$ per cent level.}
    \label{fig:comparison}
\end{figure}

In Figure \ref{fig:comparison} we show the log-derivative of the stacked subhalo density $n_s(r)$ at large scales. This is the result of a fit obtained using the model of \cite{Diemer_2014}, and we refer the reader to the aforementioned paper and \cite{Contigiani_2019b} for a detailed explanation of the model and its components. \edit{The choice to employ this profile is based on its ability} to capture the sharp feature visible around $r_\text{200m}$, which is the focus of this work. We optimally sample its $8$-dimensional parameter space using an ensemble sampler \citep{Foreman-Mackey2013}.

In the same plot, we also include the stacked subhalo profile of the accompanying dark matter only (DMO) simulations, initialized with matching initial conditions. The two profiles match almost exactly, suggesting that baryonic effects do not alter this feature to a significant extent \citep[see also][]{2020arXiv201200025O}. While not shown, we report that the same conclusion can be reached by looking at the full matter distribution $\rho(r)$ in the two sets of simulations. Similarly, this feature is also visible in the number density of galaxies, $n_g(r)$. \edit{Due to our focus on all three of these profiles, we choose not to work with background subtracted quantities.}

For reference, we present a full list of the simulated clusters used in this paper and their relevant properties, some of them defined in the following sections, in Table~\ref{tab:clusterlist}.

\begin{table*}
	\centering
	\caption{The Hydrangea clusters used in this paper and their $z=0$ properties. $\Gamma_{0.3}$ is the accretion rate measured between $z=0$ and $z=0.297$. The three splashback radii $r_\text{sp}$, $r_\text{sp}^\text{g}$, $r_\text{sp}^\text{s}$ refer to the splashback radius measured, respectively, in the dark matter, galaxy, and subhalo distributions (see Section~\ref{sec:definition}). For two clusters, CE-28 and CE-18, the radius $r_\text{sp}$ is not used in this work because the dark matter distribution displays a featureless profile at large scales. All quantities are in physical units.}
	\label{tab:clusterlist}
	\begin{tabular}{lccccccc} 
		\hline
		Name & $\Gamma_{0.3}$ & $M_\text{200m}$ & $r_\text{200m}$ & $r_\text{sp}$ & $r_\text{sp}^\text{g}$ & $r_\text{sp}^\text{sub}$\\
		 & & $[10^{14}~M_\odot]$ & [Mpc] & [Mpc] & [Mpc] & [Mpc]\\
		\hline
CE-0 & 0.8 & 1.74 & 1.74 & 2.98 & 2.72 & 2.60 \\
CE-1 & 2.0 & 1.41 & 1.63 & 1.71 & 1.56 & 1.79 \\
CE-2 & 0.5 & 1.41 & 1.63 & 2.36 & 3.27 & 2.36 \\
CE-3 & 0.8 & 2.04 & 1.84 & 2.60 & 2.72 & 2.72 \\
CE-4 & 2.8 & 2.19 & 1.89 & 1.63 & 1.87 & 1.79 \\
CE-5 & 2.0 & 2.24 & 1.90 & 2.36 & 2.60 & 2.48 \\
CE-6 & 1.1 & 3.31 & 2.16 & 2.60 & 2.48 & 2.60 \\
CE-7 & 1.2 & 3.39 & 2.17 & 3.13 & 2.60 & 2.85 \\
CE-8 & 1.8 & 3.09 & 2.12 & 2.26 & 2.48 & 2.06 \\
CE-9 & 1.1 & 4.27 & 2.36 & 3.76 & 3.76 & 3.27 \\
CE-10 & 0.8 & 3.55 & 2.21 & 3.13 & 3.13 & 2.98 \\
CE-11 & 1.4 & 4.27 & 2.34 & 3.13 & 2.85 & 2.72 \\
CE-12 & 0.1 & 5.13 & 2.49 & 3.43 & 3.76 & 4.13 \\
CE-13 & 1.5 & 5.25 & 2.52 & 2.26 & 3.13 & 2.72 \\
CE-14 & 2.1 & 6.17 & 2.66 & 2.60 & 2.72 & 2.48 \\
CE-15 & 4.2 & 6.76 & 2.73 & 1.96 & 2.26 & 2.48 \\
CE-16 & 2.7 & 7.59 & 2.84 & 1.42 & 4.13 & 3.43 \\
CE-18 & 1.1 & 9.12 & 3.03 & - & 3.76 & 3.76 \\
CE-21 & 3.7 & 12.30 & 3.34 & 2.36 & 2.85 & 2.60 \\
CE-22 & 1.5 & 16.98 & 3.72 & 4.53 & 4.53 & 4.33 \\
CE-24 & 1.5 & 15.49 & 3.61 & 3.27 & 3.27 & 4.33 \\
CE-25 & 3.4 & 19.05 & 3.87 & 3.43 & 3.43 & 3.43 \\
CE-28 & 1.9 & 21.88 & 4.06 & - & 3.94 & 3.27 \\
CE-29 & 3.5 & 32.36 & 4.61 & 3.94 & 4.13 & 3.94 \\
		\hline
	\end{tabular}
\end{table*}

\section{Splashback}
\label{sec:definition}

\subsection{Definition}

For haloes that continuously amass matter, material close to its first apocentre piles up next to the edge of the multi-stream region, where collapsed and infalling material meets \citep{Adhikari_2014}. A sudden drop in density, i.e. the feature visible in the profiles of Figure~\ref{fig:comparison}, is associated with this process. 

This intuitive picture leads to three characterizations of the splashback radius, depending on the approach used to measure or model it:
\begin{enumerate}
\item The location of the outermost phase-space caustic.
\item The point of steepest slope in the density profile.
\item The apocentre of recently accreted material.
\end{enumerate}
While these definitions have all been previously hinted at in the introduction, in this section we explicitly present them and discuss the connections existing between them. This also justifies our adopted definition, based on the density profile.
 
The first definition is clearly motivated in the spherical case but fails once it is applied to realistic haloes. The presence of angular momentum and tidal streams from disrupted subhaloes \citep[see, e.g.,][]{2011MNRAS.413.1419V}, smooth out this feature and make its description murky. The second definition was the first suggested in the literature. Introduced by \cite{Diemer_2014}, it is based on the study of dark matter profiles in N-body simulations and has been linked to the first, more dynamical, definition \citep{Adhikari_2014, Shi2016}. The third was first suggested by \cite{Diemer_2017}, who showed that this location can be calibrated to the second one \citep{Diemer_2017b} by choosing specific percentiles of the apocentre distribution. 

To clarify the relationship between the outermost caustic and apocentre, it is educational to use a self-similar toy model based on \cite{Adhikari_2014} to show the phase-space distribution of a constantly accreting halo with an NFW-like mass profile \citep{Navarro_1997}. 

In the absence of dark energy, we follow the radial motion of particles,
\begin{equation}
    \label{eq:motion}
    \ddot{r} = \frac{GM(<r, t)}{r^2},
\end{equation}
 between their first and second turnaround in the mass profile:
\begin{equation}
    \label{eq:motion2}
    M(r, t) = M(R, t) \frac{f_\text{NFW}(r/r_\text{s})}{f_\text{NFW}(R/r_s)}.
\end{equation}
We impose that the total mass evolves as $M(R, t) \propto t^{2\Gamma/3}$, $R\propto t^{2(1+\Gamma/3)/3}$, and the dimensionless NFW profile is defined as: $f(x) =\log(1+x)-x/(1+x)$. In this set of equations, $\Gamma$ is the dimensionless accretion rate, $R$ represents the turnaround radius, and the scale parameter $r_s$ is defined by the infall boundary condition 
\begin{equation}
    \label{eq:boundary}
    \frac{d\log M}{d\log r}(R) = \frac{3\Gamma}{3+\Gamma}.
\end{equation}
This condition, combined with the turnaround dynamics, imposes that $M(R, t)\propto (1+z)^{-\Gamma}$ \citep{1984ApJ...281....1F}.

We point out that the dependence on the time-sensitive turnaround properties $M(R, t), R(t)$ can be factored out from the equations above, meaning that the entire phase-space at all times can be obtained with a single numerical integration. 

\begin{figure}
    \centering
    \includegraphics[width=0.46\textwidth]{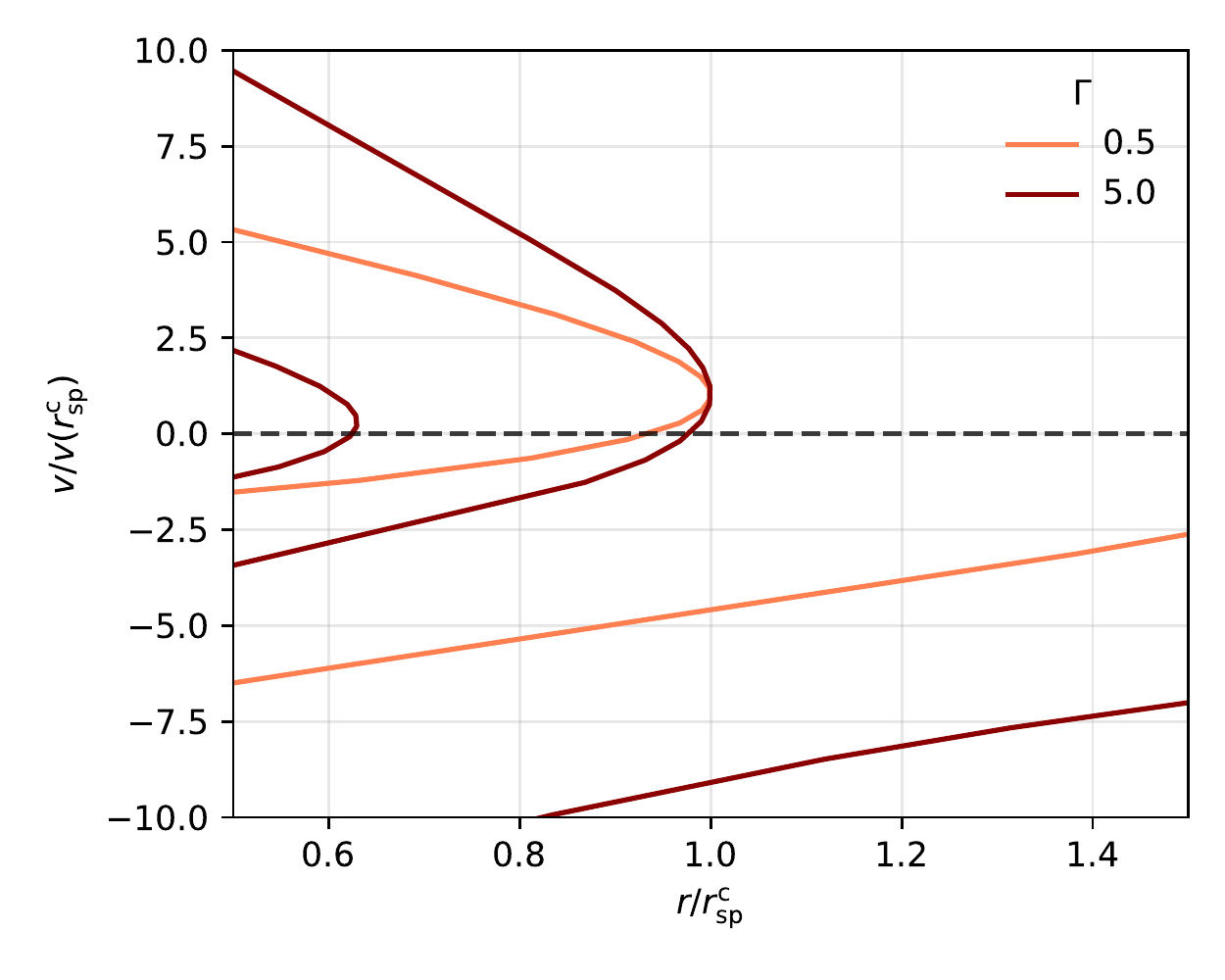}
    \caption{The phase-space structure of accreting dark matter haloes depends on the accretion rate $\Gamma$. We employ a toy model of spherical collapse to describe the multi-stream region of NFW-like haloes. The figure shows that the material at the outermost caustic, $r_\text{sp}^c$, is not necessarily at apocentre (where $v=0$) and that the ratio of these two radii is a function of accretion rate. For ease of readability, we have rescaled the coordinates by $r_\text{sp}^c$, and the velocity of collapsed material at this point. }
    \label{fig:gammavel}
\end{figure}

In Figure~\ref{fig:gammavel} we show the result of this calculation, denoting the location of the outermost caustic as $r_\text{sp}^{c}$. The caustic is formed by the outermost radius at which shells at different velocities meet ($r/r_\text{sp}^c=1$ in the plot) and the location of shells at apocenter is defined by the intersection between the zero-velocity line and the phase-space distributions. From the figure, two things are noticeable: material at $r_\text{sp}^c$ has not reached its apocentre yet, and the ratio between these two locations depends on the accretion rate. 

It is beyond the scope of this work to quantify this dependence since it depends heavily on the mass profile inside $r_\text{sp}^{c}$. Qualitatively, however, the difference between caustic and apocentre is easy to understand once the dynamical nature of this feature is considered: the halo is growing in size, and while some material is now reaching its apocentre, mass accreted more recently has the chance to overshoot it and form the actual caustic. In a static picture, this would not be the case. 

In realistic haloes, this dependence on accretion rate is only one of many factors that biases and adds scatter to the relationship between the halo boundary and apocentres. Other factors include, e.g., non-spherical orbits and the presence of multiple accretion streams. Despite this, \cite{Diemer_2017} has shown that there is a clear link between the apocentre distribution and splashback. The percentile definition introduced there is particularly suited to theoretical investigations, but its usefulness in the very low-$\Gamma$ regime is still uncertain \citep{Mansfield_2017, xhakaj2019accurately}, and it has not been explored in the presence of modifications of gravity \citep{Adhikari_2018, Contigiani_2019}.

For this work, we define the splashback radius as the location of the steepest slope as defined by a profile fit. In Table~\ref{tab:clusterlist} we report, for each cluster, this radius measured in the distribution of galaxies, subhaloes, and total matter ($r_\text{sp}^g, r_\text{sp}^s, r_\text{sp}$). The model is a modified Einasto profile \citep{Einasto1965} with the addition of a power-law to take into account infalling material \citep{Diemer_2014}. Regarding the goodness of fit, we find that up to and around $r_{200m}$ the standard deviation of the residuals is of order $10$ per cent. On the other hand, the presence of substructure superimposed on a shallow density profile results in normalized residuals of order $50$ per cent in the outer regions.

\begin{figure}
    \centering
    \includegraphics[width=0.47\textwidth]{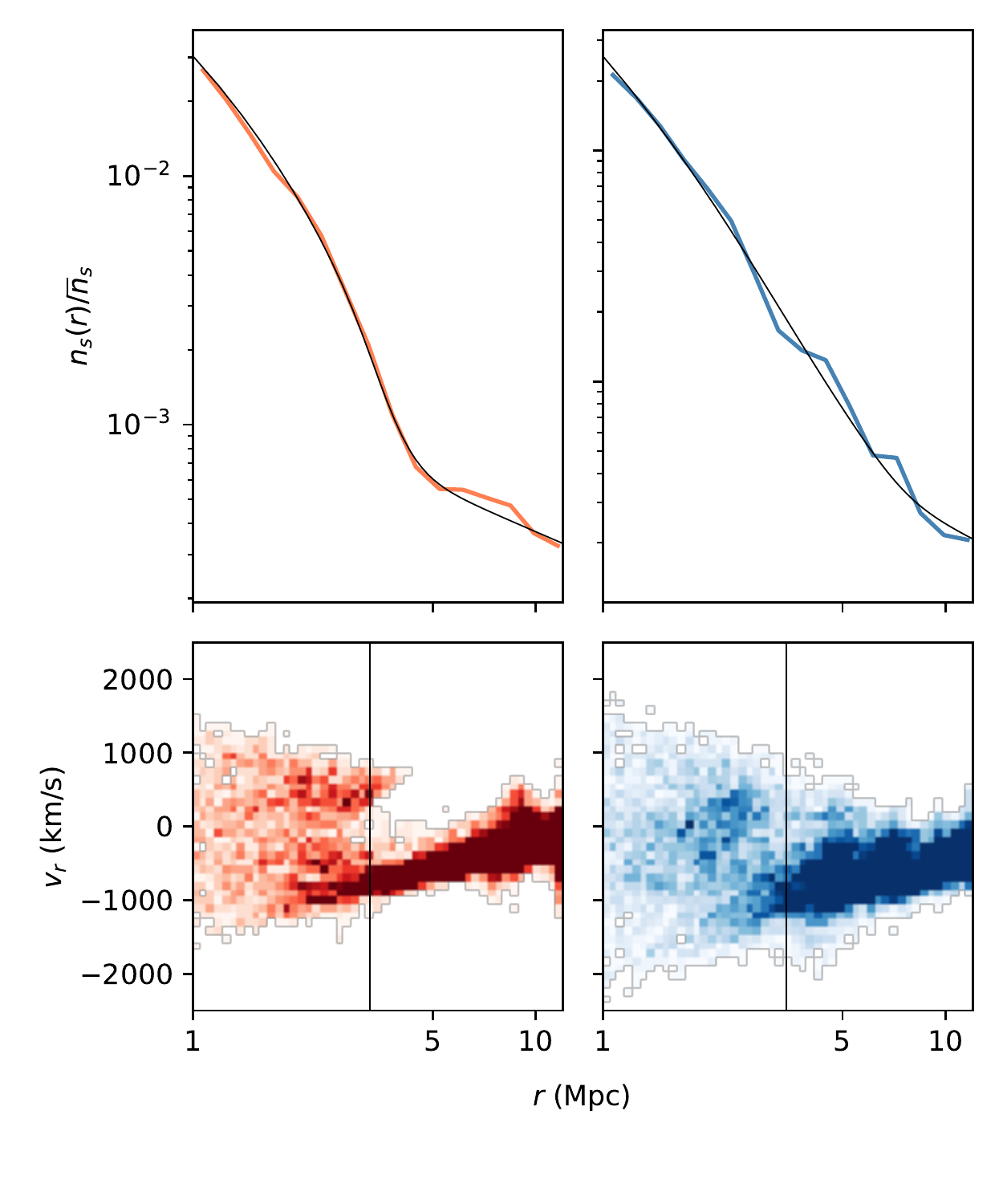}
    \caption{Fitting simulated subhalo profiles with a smooth model. In the top panels, we show the radial subhalo distributions of two clusters (CE-16, left and CE-9, right), together with the best-fit profiles used to reconstruct the log-derivative. In the bottom panels, we show how the inferred location of the log-derivative minimum (vertical line) identifies the phase-space edge of relaxed (left) and perturbed (right) galaxy clusters. In the phase-space plots, the cluster on the left is formed by collapsed particles, while the stream visible on the lower right is infalling material. The right panels demonstrate how our approach is effective even in the presence of an on-going merger when the splashback feature is not visible as a sharp transition in the density profile.}
    \label{fig:fit}
\end{figure}

To further justify our approach, we show in Figure~\ref{fig:fit} how this simple definition of splashback radius is able to capture the phase-space boundary of different haloes, even when a sudden drop in density is absent. The main benefit of this definition is that it avoids the arbitrariness of the apocentre definition, or the bias induced by multiple caustics in the minimum slope definition \citep{Mansfield_2017}. Its main caveats, however, are that 1) it is computationally expensive since it requires high-resolution simulations and a multi-parameter fit procedure, and 2) it might not apply to low-mass clusters and galaxy groups. We leave this last question open for future investigations.

We wrap this subsection up by stressing that this definition of ``the'' splashback radius is, like any other, useful only to study its correlation with other properties, or quantify the impact of different physical processes. While the flexibility of the chosen model is not surprising given the number of free parameters, the clear connection between the phase-space and the log-derivative in individual haloes is a powerful and seemingly general result. Ultimately, however, the observational results focus on stacked projected density profiles, and so should the predictions.

\begin{figure}
    \centering
    \includegraphics[width=0.46\textwidth]{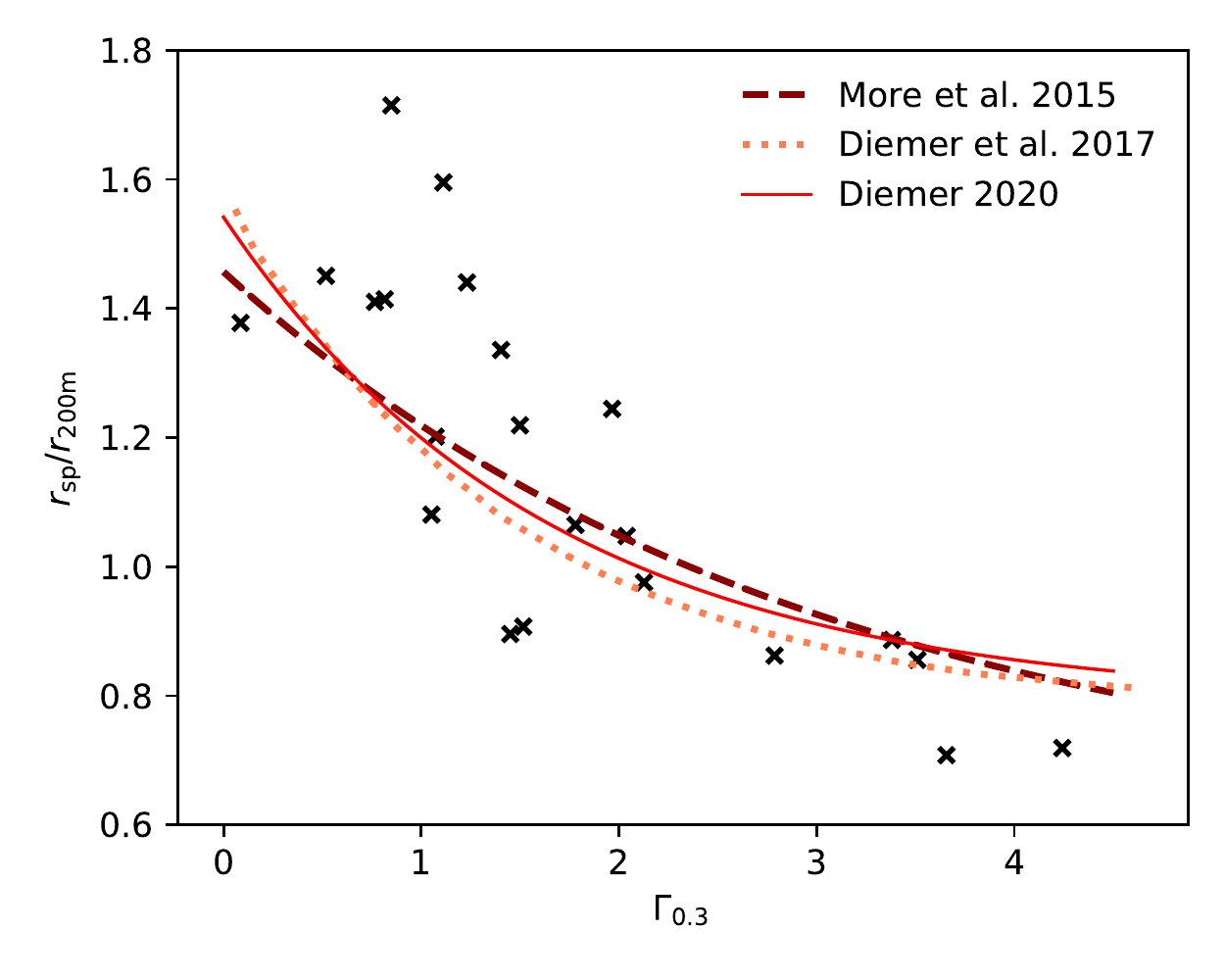}
    \caption{The splashback radius and its correlation with the accretion rate. The ratio between the splashback radius and the $200$m overdensity radius correlates with the accretion rate. We show that this correlation exists for the clusters studied in this work and compare it to the relations obtained in three other studies (see text for references). }
    \label{fig:gammarsp}
\end{figure}

\subsection{Accretion}

It is well established \citep{Diemer_2014, More_2015,Mansfield_2017,diemer2020part} that the location of the halo boundary correlates with the accretion rate
\begin{equation}
    \Gamma_{0.3} = \frac{\Delta \log M_\text{200m}}{\Delta \log (1+z)}.
\end{equation}
In this work, this ratio is calculated in the redshift range $z=0$ to $z=0.293$, since this time interval roughly corresponds to one crossing time for all clusters considered here, i.e. how long ago the material currently at splashback has been accreted \citep{Diemer_2017}. Although this choice is partially arbitrary, we have investigated the dependence of our results on the redshift upper limit and we have verified that our main conclusions are not affected. 

The archetypical relation demonstrating this idea is plotted in Figure~\ref{fig:gammarsp}, where we have also included the relations found in \cite{More_2015}, \cite{Diemer_2017b}, and \cite{diemer2020part}, to provide additional context. We find good agreement, even though a perfect match is not necessarily expected. The Hydrangea clusters represent a biased sample, selected to be mostly isolated at low redshift \citep{Bahe2017}. \edit{While the effect of this selection on the accretion rate distribution is not fully known, we} show below that a connection between cluster environment and this quantity exists, and the presence of mergers might therefore influence it. \edit{This is not surprising since a connection between accretion and large-scale bias is already known \citep[e.g., ][]{2010MNRAS.401.2245F}.}

\begin{figure}
    \centering
    \includegraphics[width=0.47\textwidth]{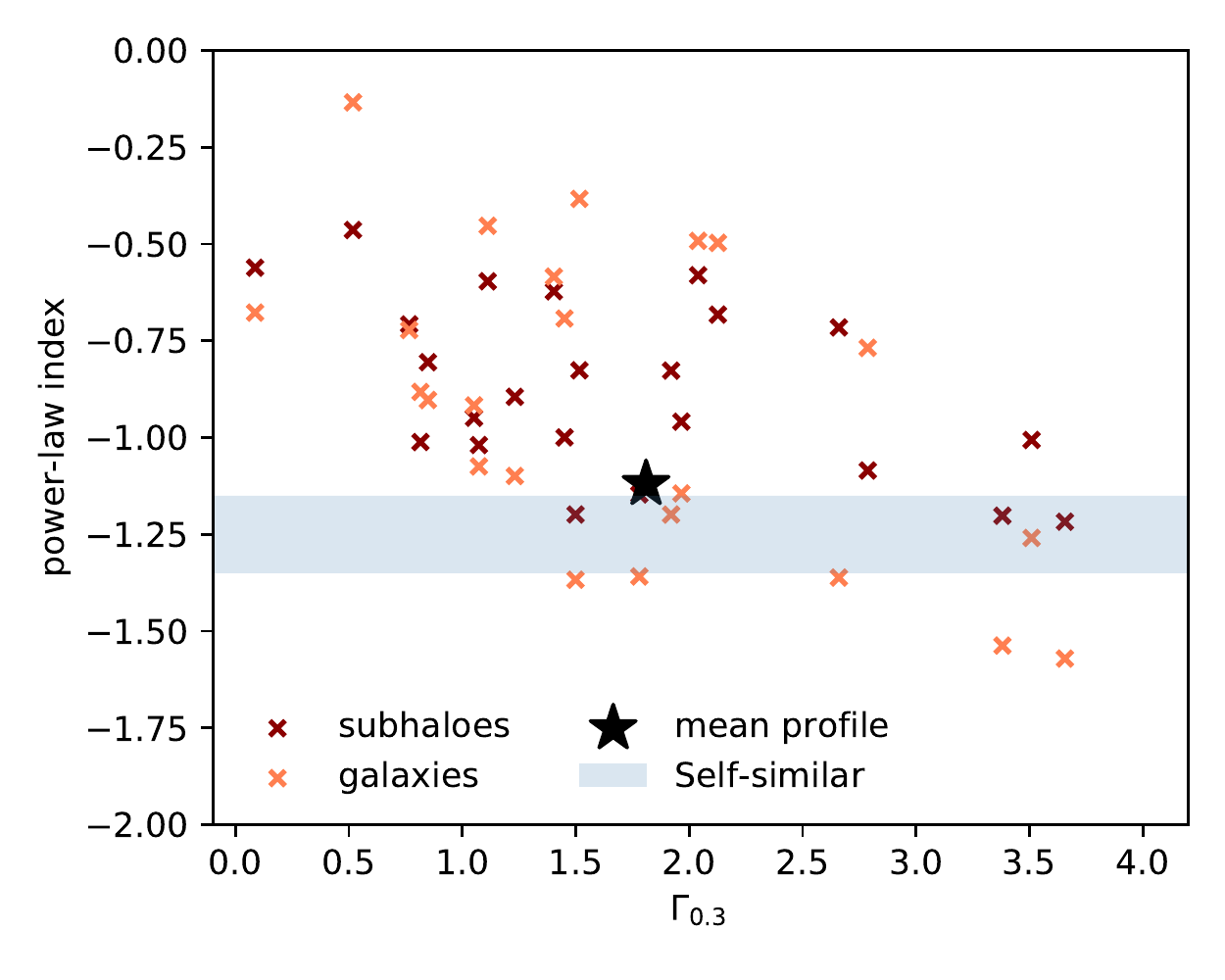}
    \caption{The distribution of subhaloes and galaxies outside the cluster edge as a function of accretion rate. Faster growing haloes display a more concentrated distribution of satellites outside of their boundary. This behaviour seen in individual clusters is not explained by simple models of spherical collapse (blue shaded area), but the average profile (marked by a star) matches the expectation. This suggests that non-isotropic processes shape this relation.}
    \label{fig:slope}
\end{figure}

We show this relationship explicitly in Figure~\ref{fig:slope} by using one of the parameters of the profile model. As visible in the figure, the power-law index of material outside of splashback correlates with the accretion rate. \edit{We find that this is true for both subhaloes and galaxies and that the difference between the two is consistent with sample variance}.

To try and explain this behaviour, we use a fully consistent model of spherical collapse introduced by \citet{1985ApJS...58...39B}, which was also used in \citet{Contigiani_2019}. The set-up of this toy model is the same as what is shown in Equation~\eqref{eq:motion}, but with a mass profile that also needs to be solved for. Starting from an initial guess for $M(r, t) = \mathcal{M}(r/R(t))$, orbits are integrated and their mass distribution is calculated. Iterating this process multiple times returns a self-similar density profile and orbits consistent with each other.

The result of this calculation is also shown in Figure~\ref{fig:slope}. Because the mass-profile prediction is not a power law, we plot a filled line displaying the range of logarithmic slopes allowed between $r_\text{sp}^\text{c}$ and $2r_\text{sp}^\text{c}$. The fact that this prediction is not a function of accretion rate implies that the correlation between the slope and the accretion rate seen in the simulations is not purely dynamical, and suggests a connection between the cluster environment and accretion rate. 

We stress here that previous splashback works have mostly focused on stacked halo profiles, for which the expectation of the spherically symmetric calculation shown above is roughly verified, even in the presence of dark energy \citep{Shi2016}. We also recover this result for our sample (see the star symbol in Figure~\ref{fig:slope}), but we point out that this is a simple conclusion. Because Newtonian gravity is additive, stacking enough clusters should always recover the spherically symmetric result. Despite this, we also note that results from the literature do not always agree with this prediction. However, we do not linger on these discrepancies since 1) this was never the focus of previous articles, and 2) different methods to extract the power-law have been employed.

\subsection{Anisotropy}

\begin{figure}
    \centering
    \includegraphics[width=0.47\textwidth]{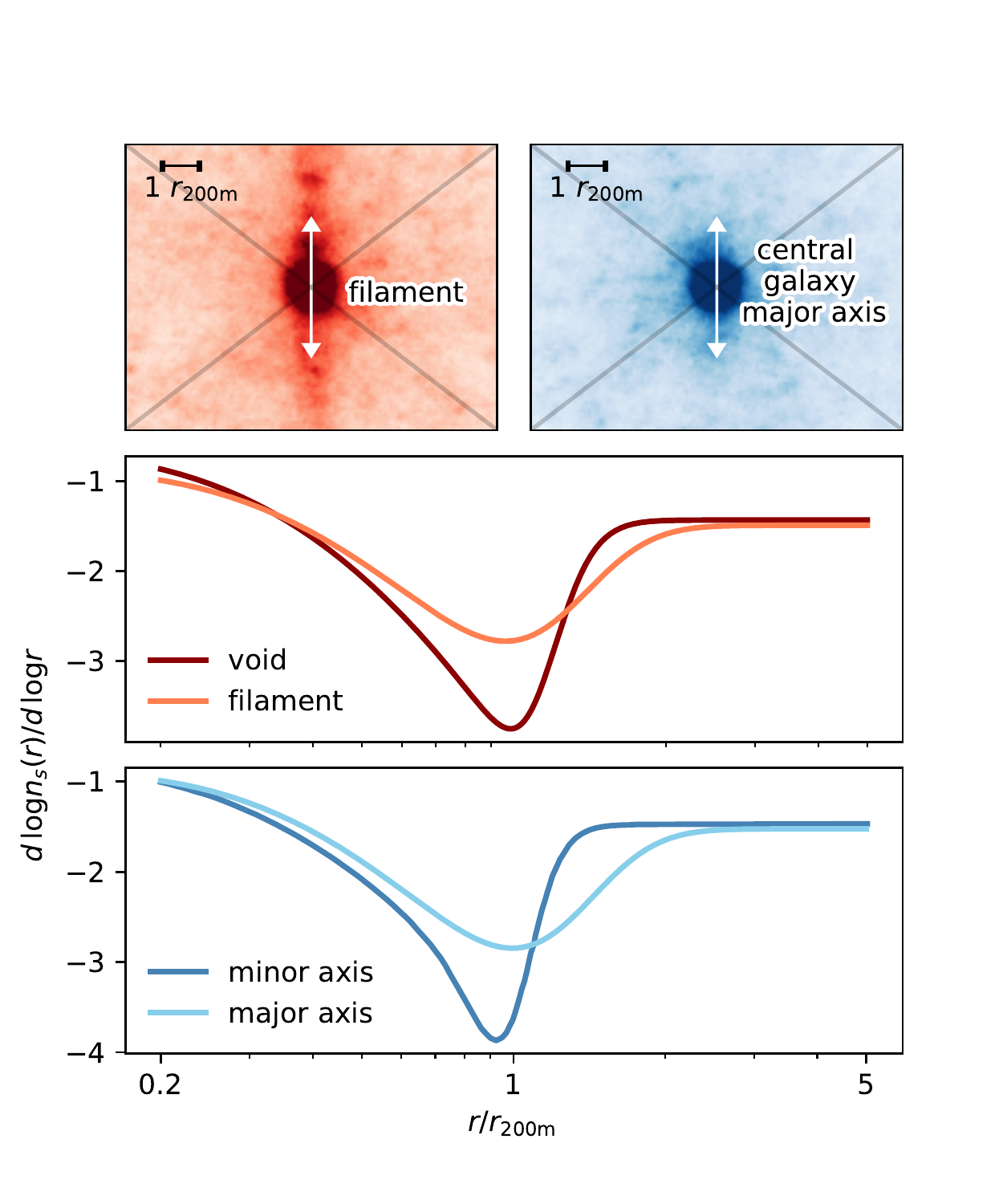}
    \caption{The impact of filaments and accretion flows on the cluster's edge. We rotate the $2$D subhalo distributions of different clusters to align their accretion axes. The top panels show the resulting mean distributions in a square region of size $5r_\text{200m}$ obtained with two definitions of this direction: one based on the presence of filaments outside $r_\text{200m}$ (left), and one based on the central galaxy's major axis (right). The first one better identifies the filamentary structures around the clusters, but the second one is closer to what can be observed. In the bottom panels, we show how the inferred $3$-dimensional logarithmic slope inside the quadrants aligned with the accretion direction (darker shade) differs from the profile outside (lighter shade). The results from the bottom panel imply that the central galaxy's major axis traces the direction of infalling material.}
    \label{fig:filaments}
\end{figure}

This departure from the spherical case implies that anisotropies play a role in shaping the accretion rate $\Gamma$. To study the impact of accretion flows on the cluster boundary, we study $72$ sky projections of the Hydrangea clusters ($3$ each, perpendicular to the $x, y,$ and $z$ axes of the simulation boxes) and rotate them to align the preferred accretion axes in these planes. For each projection, we define this direction $\theta\in (-\pi/2, \pi/2)$ in two ways: 1) to capture the filamentary structure around the cluster between $r_\text{200m}$ and $5r_\text{200m}$, we divide the subhalo distribution in $20$ azimuthal bins and mark the direction of the most populated one, and 2) to capture the major axis of the BCG, we use unweighted quadrupole moments of the central galaxy's stellar profile within $10$ kpc from its centre. The mean projected distributions according to these two methods are presented in the left and right top panels of Figure~\ref{fig:filaments}, respectively.

Looking at the top-left panel of the figure, it is not surprising that filamentary structures of the cosmic web are visible around the central cluster -- this is by construction. \edit{Because of the higher contrast between outside and inside regions, the subhalo distribution exhibits a sharper feature in the directions pointing towards voids (see central panel of Figure~\ref{fig:filaments}).} More surprisingly, however, these same traits are also noticeable in the mean distributions aligned according to the central galaxy's axis (see lower panel).

This result implies that the distribution of stellar mass within the central $10$ kpc of the cluster contains information about the distribution of matter at radii which are a factor $10^2$ larger. In fact, the connection between the shape of the dark matter halo and the ellipticity of the brightest cluster galaxy (BCG, which is also the central galaxy for massive galaxy clusters) is known \citep{Okumura_2009, Herbonnet_2019, 2020MNRAS.495.2436R}. And, similarly to other results \citep{Conroy_2007, 2007MNRAS.375....2D}, the Hydrangea simulations predict that the stellar-mass buildup of the BCG is driven by the stripping of a few massive satellites after their first few pericentre passages (Bah\'{e} et al., in prep.). Because these galaxies quickly sink to the centre, the material they leave behind is, therefore, a tracer of their infalling direction.

\section{The mass--size relation}
\label{sec:msrelation}

In our sample, we find that the splashback feature seen in the galaxy, subhalo, and total matter profiles are all at the same location. The mean fractional difference between any two of $r_\text{sp}^g, r_\text{sp}^s$, or $r_\text{sp}$ is consistent with zero, with a mean standard deviation of $3$ per cent. We also verified that this statement is unaffected by cuts in subhalo mass or galaxy stellar mass. \edit{Due to the limited size of our sample, the effects of dynamical friction on the distribution of high-mass subhaloes are not visible \citep{2016JCAP...07..022A}.} 

We emphasize, however that this does not mean that galaxy selection effects have no impact on these quantities. For example, it is an established result, both in the Hydrangea simulations \citep{oman2020homogeneous} and in observations \citep{2020arXiv200811663A}, that the location of a galaxy in projected phase-space correlates with its colour and star-formation rate. This is because a red colour preferentially selects quenched galaxies that have been orbiting the halo for a longer time. 

Until their first apocentre after turnaround, galaxies act as test particles orbiting the overdensity as the halo grows in mass. In the standard cold dark matter paradigm, based on a non-interacting particle, it is not surprising then that the edge formed in their distribution is identical to the one seen in the dark matter profile. It should be noted, however, that this is not necessarily true in extended models in which dark matter does not act as a collisionless fluid.  Due to their infalling trajectories, the distribution of galaxies will always display a splashback feature, even if the dark matter profile does not exhibit one. 

In the cold dark matter scenario, our result implies that galaxies can be used to trace the edge of clusters. We note, in particular, that this measurement has already been performed several times using photometric surveys \citep{Baxter_2017, Nishizawa_2017, Chang_2018, Z_rcher_2019, Shin_2019}. Furthermore, due to the large number of objects detected, galaxy distributions obtained through this method offer the most precise measurements of splashback. The accuracy of the results, however, depends heavily on the details of the cluster finding algorithm \citep{Busch_2017, Shin_2019}. 

\begin{figure}
    \centering
    \includegraphics[width=0.47\textwidth]{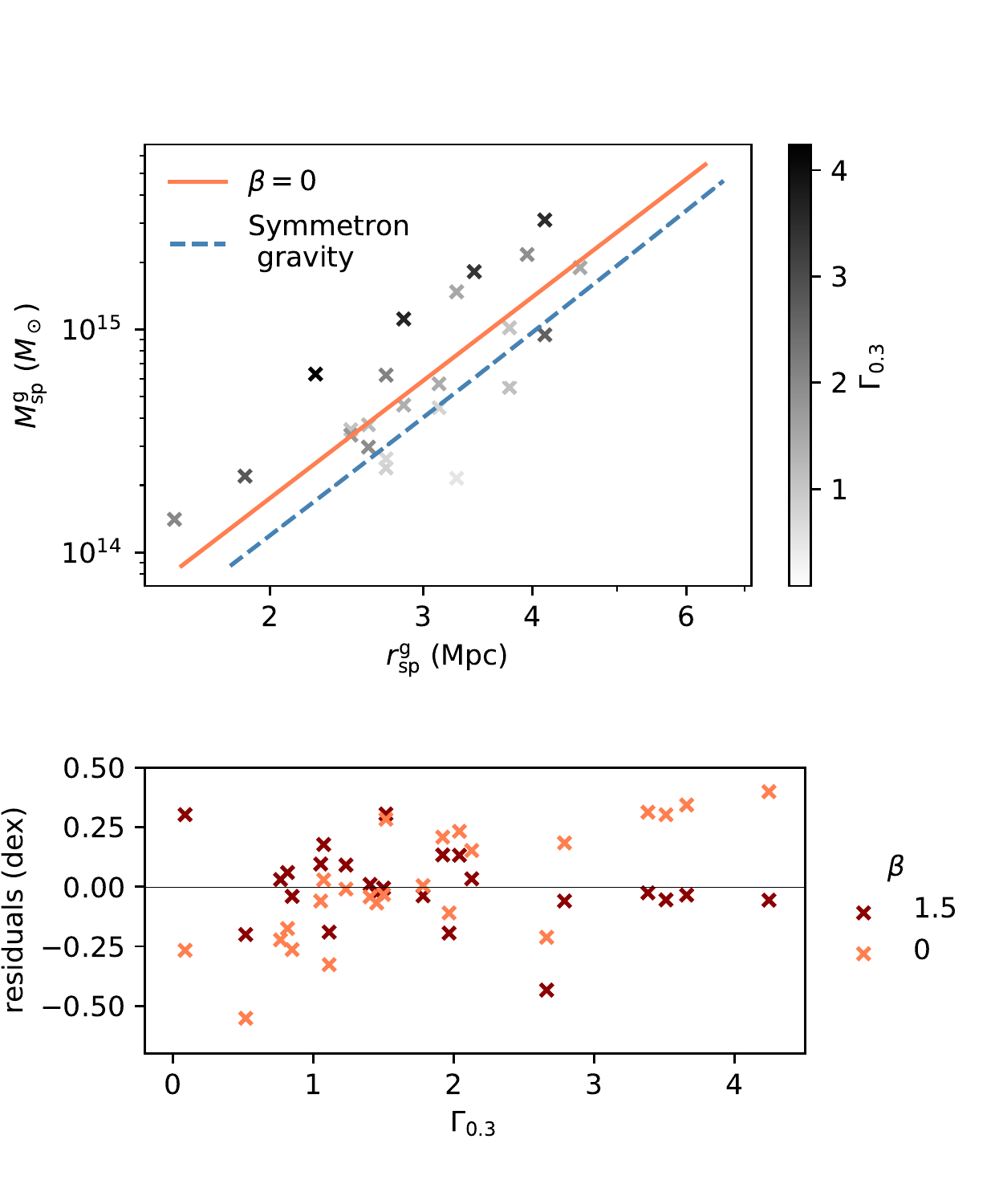}
    \caption{ The mass size relation of galaxy clusters. In the top panel, we show how the size of the cluster boundary seen in the galaxy distribution, $r_\text{sp}^{g}$, scales with its enclosed mass, $M_\text{sp}^g$. In the same panel we also show the median relation in Equation~\eqref{eq:scaling} obtained for $\beta = 0$ and how modifications of gravity are expected to affect this relation (blue dashed line, see text for more details). In this relation, a secondary dependence on the accretion rate $\Gamma_{0.3}$ is a source of scatter that can be captured if $\beta\neq 0$. As visible in the residuals in the bottom panel, a simple power-law form well reproduces this dependence. In the considered sample, we find that half of the total scatter ($0.25$ dex) is due to the mass accretion rate distribution.}
    \label{fig:msscaling}
\end{figure}

With this in mind, we build an observational mass-size relation between the location of this feature in the galaxy distribution ($r_\text{sp}^g$) and the mass enclosed within it ($M_\text{sp}^g$). In Figure~\ref{fig:msscaling} we present the correlation between the two for the Hydrangea clusters. Because the splashback radius is roughly located at $r_\text{200m}$ (see Figure~\ref{fig:gammarsp}), this relationship can be understood as a generalization of the virial mass-radius relation, where we have introduced a dependence on accretion rate. Surprisingly, we find that the dependence on $\Gamma_{0.3}$ is well captured by a simple form:
\begin{equation}
    \label{eq:scaling}
    \frac{M_\text{sp}}{r_\text{sp}^3 } \propto (1+\Gamma_\text{0.3})^{\beta}.
\end{equation}
While we do not constrain $\beta$ precisely, we find that $\beta = 1.5$ provides an adequate fit by reducing the total scatter from $0.25$ dex to about half of this value. This choice of exponent and functional form is supported by the model of self-similar collapse used for Figure~\ref{fig:slope}, where we find that a power-law $\beta=1.45$ fits this relationship with the same precision as the exponential functions calibrated to numerical simulations shown in Figure~\ref{fig:gammarsp}. For a more extensive comparison with these predictions, we refer the reader to Section~\ref{sec:conclusions}.

The virial relation is a trivial connection between the mass and size of haloes based on an overdensity factor, but its observational power is limited by the fact that these masses are usually extracted from parametric fits to weak-lensing profiles that do not extend to the respective overdensity radii. Because of this, the overdensity masses have a strong dependence on the assumed mass-concentration relation \citep[see, e.g.][]{Umetsu_2020}. The splashback feature, on the other hand, naturally predicts a mass-size relation for galaxy clusters and does so without the need for external calibrations. 

In Figure~\ref{fig:msscaling} we also plot the expected change in this relation to due modifications of gravity. We use the symmetron gravity model of \cite{Contigiani_2019} with parameters $f=1$ and $z_\mathrm{ssb}=1.5$, and assume that the change affects only the splashback radius and not the mass contained within it. The exact result depends on the theory parameters, but the expected change in this relation is around $0.15$ dex.

Experimentally, we argue that this relation can be probed using a combination of galaxy density profiles (to extract $r_\text{sp}^g$) and weak lensing measurements. Aperture masses \citep{2000ApJ...539..540C}, in particular, can be used to extract in a model-independent fashion the average projected mass within a large enough radius. If necessary, the aperture mass can also be deprojected to obtain a low-bias estimate \citep{Herbonnet_2020}.

\section{Redshift evolution}
\label{sec:redshift}
So far, we have only considered the simulation predictions at $z=0$. In this section, we extend our analysis to higher redshifts by exploring two other snapshots of the Hydrangea simulations at $z=0.474$ and $z=1.017$. 

\begin{figure}
    \centering
    \includegraphics[width=0.46\textwidth]{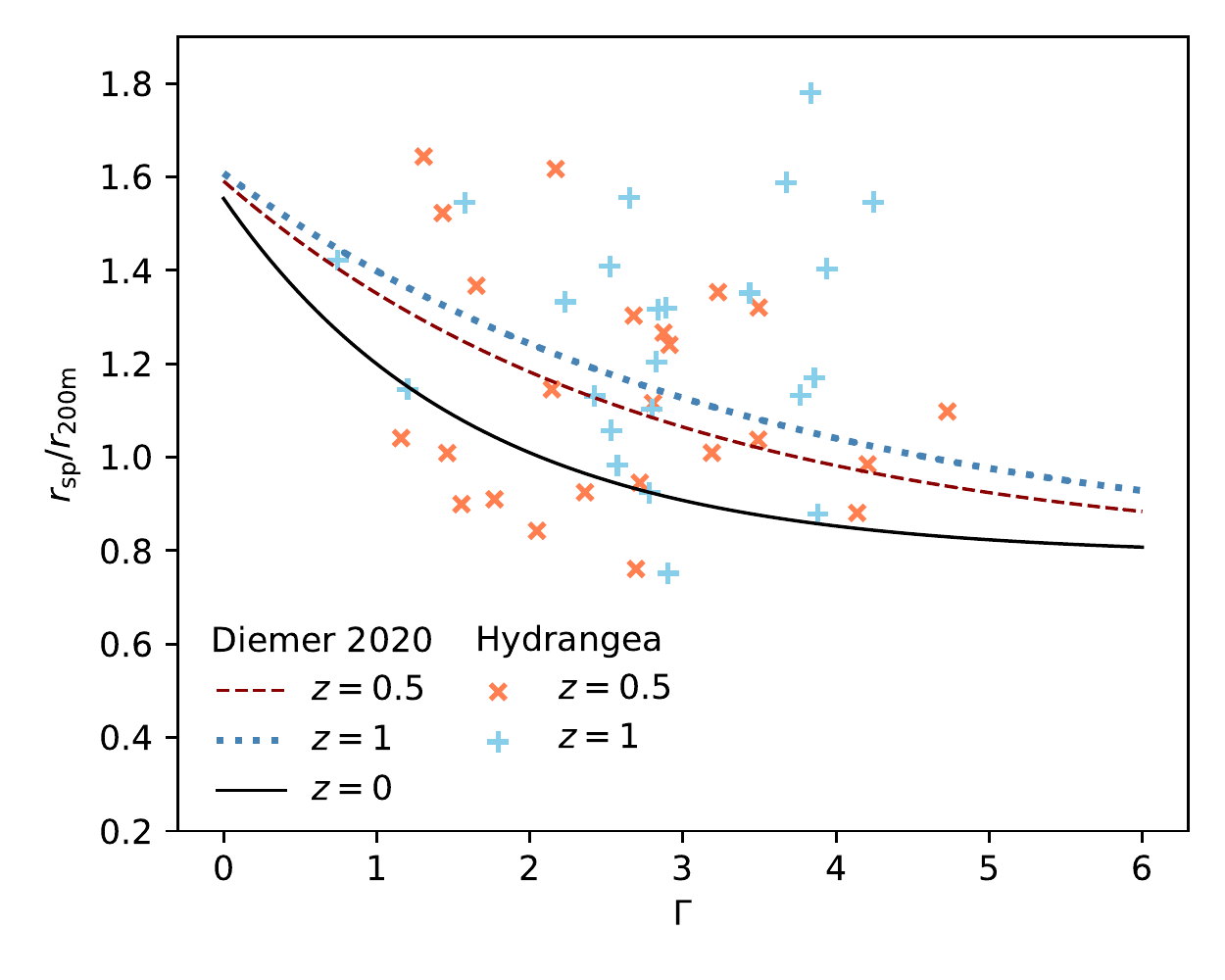}
    \caption{The splashback radius and its correlation with the accretion rate as a function of redshift. This plot is an extension of Figure~\ref{fig:gammarsp} for redshifts $z=0.474$ (orange crosses) and $z=1.017$ (light blue plus symbols). The ratio between the splashback radius and the $200$m overdensity radius should correlate with the accretion rate $\Gamma$, but for the Hydrangea snapshot at $z=1.017$ the large sample variance washes out this correlation. Despite this, we still recover the expectation of previous results (plotted lines), a larger average $r_\mathrm{sp}/r_\mathrm{200m}$ at higher redshift. }
    \label{fig:gammarspz}
\end{figure}

At these higher redshifts, we find that the scatter in the splashback relation for individual haloes is large. This is visible in Figure~\ref{fig:gammarspz}, where we plot the equivalent of Figure~\ref{fig:gammarsp} for these two snapshots. We recover the general result of \cite{diemer2020part} that the average values of $r_\mathrm{sp}/r_\mathrm{200m}$ and $\Gamma$ should be higher at early times, but the correlation between the two is completely washed out by $z=1$. We connect this to three causes: 1) The fixed time interval between the snapshots does not allow us to reliably estimate $\Gamma$ at higher redshift when the crossing times are smaller. 2) The lower number of resolved galaxies and subhaloes means that the residuals of the individual profile fits are larger around the virial radius. And finally, 3) the higher frequency of mergers at high redshift means that the numbers of haloes with profiles not displaying a clear splashback feature increases. 

We find that Equation~\ref{eq:scaling} is still valid, even if our ability to constrain the scatter at high redshift is impeded by the sample variance. Furthermore, we report that the splashback overdensity $M_\text{sp}/r_\text{sp}^3$ has a redshift dependence. Or, in other words, that the logarithmic zero-point that was not specified in Equation~\ref{eq:scaling} is a function of redshift. Not accounting for the $\Gamma$ dependence, our best fit values for the logarithm of the average overdensity $\log_{10}(M_\text{sp}/\text{M}_\odot)-3\log_{10}( r_\text{sp}/\text{Mpc})$ are $[13.3, 13.8, 14.1]\pm 0.3$ at redshifts $[0, 0.5, 1]$.

\begin{figure}
    \centering
    \includegraphics[width=0.46\textwidth]{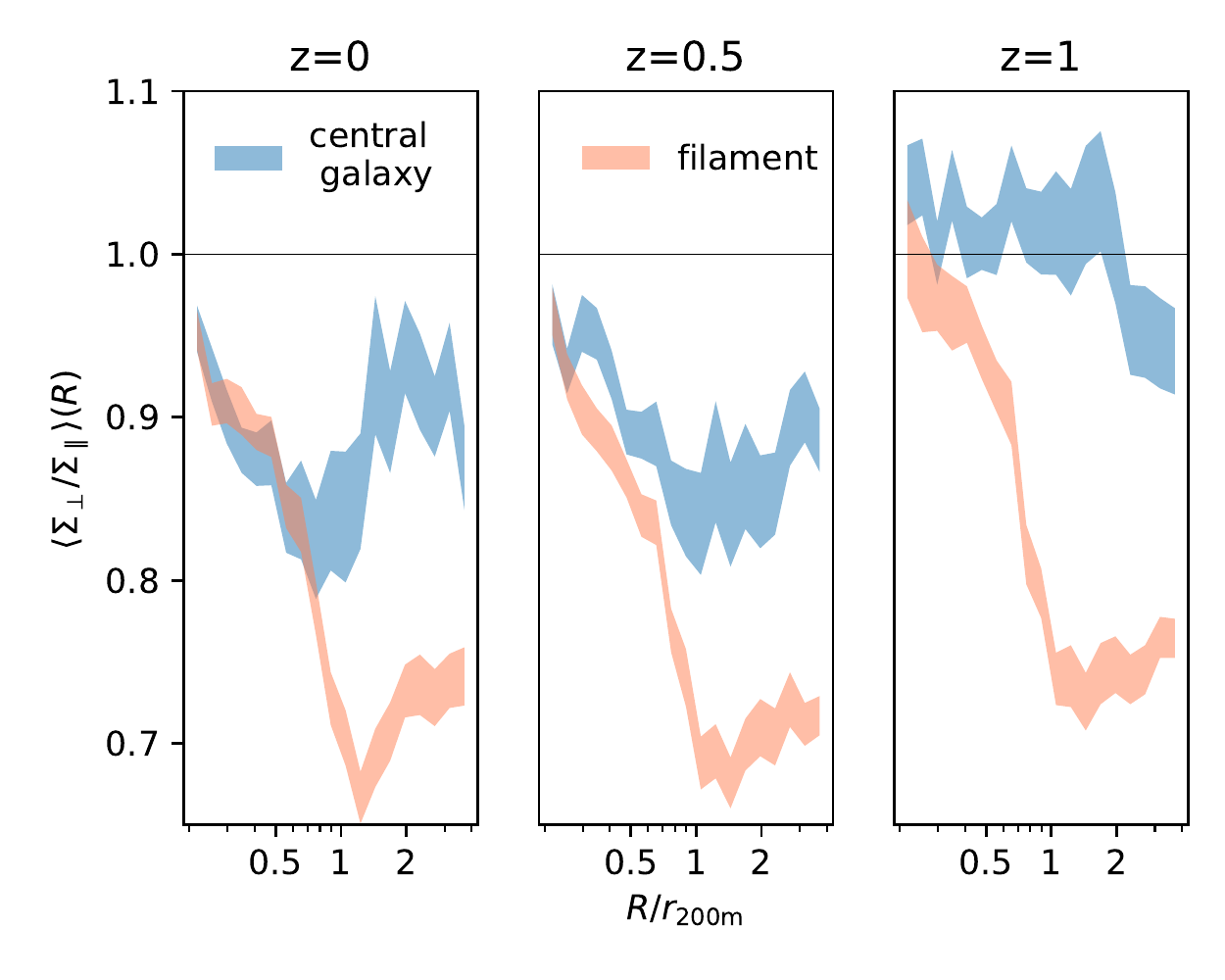}
    \caption{The impact of filaments and accretion flows on the outer density profile of massive haloes as a function of redshift. We plot the mean value and variance of the ratio between the $2$D subhalo distributions in quadrants perpendicular ($\Sigma_\perp$) and parallel ($\Sigma_\parallel$) to the accretion direction defined through two tracers. This ratio is closely related to what can be measured in observations. While the difference in profile towards and away from filamentary structures is visible at all redshifts, the orientation of the central galaxy is not a good tracer of the splashback anisotropy at high redshift. This is because the central BCG is still forming and its orientation is not yet finalized.}
    \label{fig:filamentsz}
\end{figure}

Regarding the anisotropy in the splashback feature due to filamentary structures, we report that this phenomenon exists also at high redshift. In Figure~\ref{fig:filamentsz} we compare the sky-projected subhalo profiles $\Sigma_s(R)$ towards different directions, similarly to what we have done for Figure~\ref{fig:filaments}. In this case, however, we explicitly discuss the connection with observations by plotting directly the ratio of the density profiles inside quadrants oriented towards and perpendicular to the two accretion directions, instead of focusing on the result of the profile fits. The mean and variance of these ratios are calculated assuming that the different projections are independent. We find that the orientation of the major axis of the brightest cluster galaxy does not correlate with a splashback anisotropy at $z=1$.  This is because, in most cases, the identification of a central, brightest galaxy is not straightforward at this redshift. At early times, the future central galaxy is still in the process of being created from the mergers of multiple bright satellites located close to the cluster's centre of potential. 

To conclude this section, we point out that in the region around $r_\mathrm{200m}$, the difference between the profiles perpendicular and parallel to the central galaxy's major axis is about $10$ per cent at redshift $z\lesssim0.5$. This departure is well within the precision of galaxy profiles extracted from large surveys \cite[e.g.][]{2020arXiv200811663A}. Therefore, this measurement might already be possible using such catalogues.

\section{Discussion and conclusions}
\label{sec:conclusions}

On its largest scales, the cosmic web of the Universe is not formed by isolated objects, but by continuously flowing matter distributed in sheets, filaments and nodes. For accreting (and hence non-virialized) structures such as galaxy clusters, the splashback radius $r_
\text{sp}$ represents a physical boundary motivated by their phase-space distributions. \edit{To exploit the information content of this feature, in this paper we have introduced and studied two observable quantities related to it. }

\edit{First, we have shown that the full galaxy profile can be used to define a cluster mass, i.e. the mass within $r_\mathrm{sp}$. This is an extension of the traditional approach of using richness as mass proxy \citep[see, e.g., ][]{Simet_2016}.} Because of the dynamical nature of the equivalent feature in the dark matter profile, we conclude that, observationally, the splashback feature in the galaxy profile \emph{defines} the physical halo mass. Moreover, we have shown here that the natural relation between the mass and size of haloes according to this definition (see Fig.~\ref{fig:msscaling}) can be used to constrain new physics at cluster scales. Because this boundary is delimited by recently accreted material, we found that a majority of the scatter in this mass-size relation can be explained through a secondary dependence on accretion rate $\Gamma$. 

\edit{Second, we have explored how this connection to the accretion rate might be interpreted as a connection between the geometry of the cosmic web and how clusters are embedded in it.} The relation between the two is made explicit in Figure~\ref{fig:slope}, Figure~\ref{fig:filaments}, and Figure~\ref{fig:filamentsz}. In these figures, we have investigated how the cluster environment affects both the halo growth and the stellar distribution of the central galaxy. This information, combined with the scatter of the mass-size relation, can therefore be used as a consistency check for any property that claims to select for accretion rate. 

\subsection{The role of simulations}
\label{sec:sims}

In the last few years, the study of the splashback feature has evolved into a mature field both observationally and theoretically. We use this section to discuss explicitly the connection between the two, in light of this work and its connection to previous endeavours.

\edit{In the context of splashback, simulations have guided the formulation of theoretical principles and hypotheses. However, as more measurements become viable, it becomes necessary to provide clear and powerful observables.} Following this spirit, we used high-resolution hydrodynamical simulations to explore directly the connection between measurements based on sky-projected galaxy distributions and theoretical predictions. 

Our conclusions regarding the mass-size relation and its redshift evolution are similar to the results of \cite{Diemer_2017b} and \cite{diemer2020part}, which are based on more extensive N-body simulations. For the sake of completeness, it important to note that, in the same papers, it was also found that the splashback overdensity is not universal, but has both a mild dependency on $M_\mathrm{200m}$ and a strong dependency on cosmology, especially at low redshift ($z\approx0.2$). Due to our limited sample, we are clearly unable to model these effects in this work. Nonetheless, we point out that our goal here is to construct a pure splashback scaling-relation based on galaxy profiles and weak lensing mass measurements. Every other dependency, if present, should be captured either as additional scatter or through different parameter values.

We also point out that that these previous works are based on the apocentre definition of splashback (see Section~\ref{sec:definition}). In contrast, we defined the splashback radius as the point of steepest slope according to a model fit to the density profiles of galaxy clusters. While we do not necessarily expect the two definitions to differ, our choice is based on its connection to observations, and the desire to highlight the fact that the splashback radius is not only some abstract halo property but can be defined as a characteristic of individual profiles, such as, e.g., the concentration parameter \citep{Navarro_1997}. 

\edit{An alternative method employed by other studies \citep{Diemer_2014, Mansfield_2017, xhakaj2019accurately} to obtain a measure of $r_\mathrm{sp}$ makes use of the minimum of the logarithmic slope in smoothed profiles. While this approach is much faster than profile fitting when only $r_\mathrm{sp}$ is of interest, it does not describe the full shape visible in Figure~\ref{fig:comparison}. In particular, a model that captures the width of this feature is necessary to define the slope of the outer region without an arbitrary choice of which radial scales to consider. Because the model used here contains an asymptotic outer slope, this definition is unique.}

Our decision has, of course, its drawbacks. The versatility of the fitted model is necessary to capture the variance of the individual profiles, but the resulting intrinsic scatter is large and not the best suited to study tight splashback correlations (such as Figure~\ref{fig:gammarsp}). At the same time, the large parameter space might also be seen by some as a chance to study a multitude of correlations between different model parameters. However, we resist this temptation, as inferences based on such correlations might say more about the particular model employed than provide any physical information.

A subtler difference between our method to characterize splashback and other ones present in the literature is related to the definition of spherical density profiles. \cite{Mansfield_2017} and \cite{2020MNRAS.tmp.3386D} found that the most successful method to achieve a clear splashback feature for individual haloes is to measure the median profile along multiple angular directions. In light of the results of Figure~\ref{fig:filaments}, we argue that the distribution of splashback as a function of direction is skewed by the presence of a few dense filaments and hence the difference between a median and mean splashback can be substantial. Therefore, we stress that future works should exercise caution when employing such methods. The use of median profiles smooths substructure by focusing on the halo boundary in the proximity of voids, but because this process is itself correlated with the halo growth rates (see Figure~\ref{fig:slope}), the connection with observations is not as simple as one might expect.

\subsection{Next steps}

Because in this work we have focused only on high-mass objects ($M\sim10^{14.5}~M_\odot$), a natural future step is to investigate if the results apply also in other regimes. For example, a larger sample over a wide range of masses and redshifts is required to confirm the simple form of Equation~\eqref{eq:scaling} and verify if it applies to lower mass groups ($M\sim10^{13.5}~M_\odot$).

Exploring a wider range in mass, both in observations and simulations, can also be used to confirm a key prediction: because the median accretion rate is expected to be a function of mass and redshift \citep{More_2015}, we expect the mass-size relation for an observed halo sample to not necessarily follow a simple form.

Finally, we point out that our results encourage a concentrated effort towards understanding the relationship between cluster environment and splashback. What is discussed in Figure~\ref{fig:slope}, \ref{fig:filaments} and \ref{fig:filamentsz} suggests that the connection between accretion-flows, filaments, and cluster boundary is not a simple one. To better understand this process, it will become necessary to complement the usual \emph{inside-out} theoretical approaches to splashback, that look at haloes to define their boundaries, with \emph{outside-in} approaches, that connect the cosmic web to its nodes. In this context, the amount of splashback data gathered by projects such as the Kilo Degree Survey \citep{2013ExA....35...25D}, Dark Energy Survey \citep{2005astro.ph.10346T}, and, in the future, LSST \citep{2009arXiv0912.0201L} and \emph{Euclid} \citep{laureijs2011euclid}  will provide a powerful probe for the study of structure formation.

\section*{Acknowledgements}
We thank Benedikt Diemer for providing valuable feedback on the manuscript. 
OC is supported by a de Sitter Fellowship of the Netherlands Organization for Scientific Research (NWO). HH acknowledges support from the VICI grant
639.043.512 from NWO. YMB acknowledges funding from the EU Horizon 2020 research and innovation programme under Marie Sk{\l}odowska-Curie grant agreement 747645 (ClusterGal) and the NWO through VENI grant 639.041.751. The Hydrangea simulations were in part performed on the German federal maximum performance computer ``HazelHen'' at the maximum performance computing centre Stuttgart (HLRS), under project GCS-HYDA / ID 44067 financed through the large-scale project ``Hydrangea'' of the Gauss Center for Supercomputing. Further simulations were performed at the Max Planck Computing and Data Facility in Garching, Germany. 

\bibliographystyle{mnras}
\bibliography{bibliography}

\bsp	
\label{lastpage}
\end{document}